\documentclass[%
 reprint,
 aps,
 pre,
 showkeys,
 preprintnumbers
]{revtex4-2}
\usepackage[english]{babel}

\usepackage{graphicx}
\usepackage{dcolumn}
\usepackage{bm}
\usepackage{siunitx}  
\usepackage{amsmath}
\usepackage{tabularx}
\usepackage{subfigure}
\usepackage[normalem]{ulem}
\usepackage[hidelinks]{hyperref}
\usepackage{siunitx,booktabs}
\usepackage[margin=1in]{geometry} 
\usepackage{cancel}

\usepackage{array}
\newcolumntype{P}[1]{>{\centering\arraybackslash}p{#1}}

\newcommand{\p}[1][]{\varphi_{#1}}
\newcommand{\pdc}[1][]{\varphi_{{#1}dc}}
\newcommand{\pxdc}[1][]{\varphi_{{#1}xdc}}
\newcommand{\px}[1][]{\varphi_{{#1}x}}

\begin{document}

\topmargin=-0.45in
\evensidemargin=0in
\oddsidemargin=0in
\textwidth=6.5in

\textheight=9.0in
\headsep=0.25in 

\title{Nonequilibrium Thermodynamics of a Superconducting Szilard Engine}

\author{Kuen Wai Tang}
\email{tkwtang@ucdavis.edu}
\author{Kyle J. Ray}
\email{kjray@ucdavis.edu}
\author{James P. Crutchfield}
\email{chaos@ucdavis.edu}
\affiliation{Department of Physics and Astronomy, University of California-Davis, One Shields Avenue, Davis, CA 95616}

\begin{abstract}
We implement a Szilard engine using a $2$-bit logical unit consisting of inductively coupled quantum flux parametrons (QFPs)---Josephson-junction superconducting circuits with applications in both the classical and quantum information processing regimes. Detailed simulations show that it is highly thermodynamically efficient while functioning as a Maxwell demon---converting heat to work. The physically-calibrated design is targeted to direct experimental exploration. However, variations in Josephson junction fabrication introduce asymmetries that result in energy inefficiency and low operational fidelity. We provide a design solution that mitigates these practical challenges. The resulting platform is ideally suited to probe the thermodynamic foundations of information processing devices far from equilibrium.
\end{abstract}

\date{\today}


\preprint{arxiv.org:2407.20418}

\maketitle

 

\section{Introduction}
\label{sec:Introduction}

In 1857, physicist James Clerk Maxwell conjured a thought experiment that called into question the very foundations of thermodynamics \cite{Maxwell_2011}. He envisioned an intelligent agent, capable of discerning the speed of individual gas molecules, stationed at a gate separating two compartments. With each passing molecule, the agent would deftly open or close the gate, meticulously sorting the fast-moving molecules into one compartment and the slow-moving ones into the other. Thus, through, observation and control, this clever manipulation of thermal energy, Maxwell argued, would lead to a violation of the Kelvin-Planck statement of the Second Law of Thermodynamics: No process exists whose sole result is to extract energy from a heat bath and convert it all to work.

In 1929, Leo Szilard resolved the paradox of, what came to be called, Maxwell's demon \cite{Szilard1964OnTD}. Szilard refined Maxwell's demon  by using a purely physical system---a memory---to replace the intelligent agent. Using a construction now commonly called a \emph{Szilard engine}, he argued that if the Second Law was to hold, the measurement process of storing and retrieving information from this memory must produce a compensating amount of entropy. Later, in 1961, Landauer \cite{Landauer_1961} showed that such memory manipulations must incur a cost commensurate with the Second Law. This ``no free lunch'' trade-off is most clearly illustrated when the memory system is put on an even playing field with the system being controlled. From this perspective, it is seen that the physical intervention of measurement is the inverse of the control step in which memory is leveraged to extract work. And, thus, any gain from one step is offset by a cost in the other \cite{PhysRevLett.116.190601,ray2020variations}. 

To explore the thermodynamics of information processing, the following implements a Szilard engine consisting of two coupled quantum flux parametrons (QFPs) \cite{PhysRevB.46.6338, ray2022gigahertz, arXiv:2307.01926v3, PhysRevResearch.3.033115, PhysRevB.46.6338, PhysRevLett.75.1614, Takeuchi_2013, hosoya1991quantum, 1487397}, a device arising from the intersection of superconducting quantum interference device (SQUID) and information processing literatures. This type of circuit is also known as a \emph{variable $\beta$ rf-SQUID} \cite{variable_beta, PhysRevLett.63.1712}, a \emph{gradiometric flux qubit} \cite{PhysRevResearch.2.013249}, a \emph{parametric quantron} \cite{1059351, likharev1982classical, 1063673}, and likely other names as well. Here, we choose to call them QFPs. While such SQUID-based circuits have applications in both classical and quantum information processing \cite{Berkley_2010, friedman2000quantum}, our focus is on classical information processing. Notably, the substrate and control protocol do not require continuous feedback monitoring for the engine to operate. This allows well-founded measurements of the minimal information costs associated with a Szilard engine throughout its operation.

A central motivation for using superconducting technology is that at low temperatures most heat flow pathways freeze-out. This allows the circuit to be used for direct validations of Landauer's Principle, which requires independent measurements of the nonlinear dynamics that supports information processing, on the one hand, and of thermodynamic entropy production, on the other. To emphasize, the superconducting aspect of the implementation is not for improving thermodynamic efficiency from using zero-resistance supercurrents. Indeed, our Szilard engine design can be implemented in any physical substrate that is dynamically equivalent, including room temperature electronics. The Josephson effect is used only to provide the nonlinearity needed for information storage and control, while the low temperature operation anticipates independent calorimetric measurements for isolating the heat flow implied by Landauer's principle.

Section II reviews Szilard's original engine,  establishing the theoretical relationships between information erasure and creation, as well as work costs and gains.

\begin{figure*}[ht!]
\includegraphics[scale=0.5]{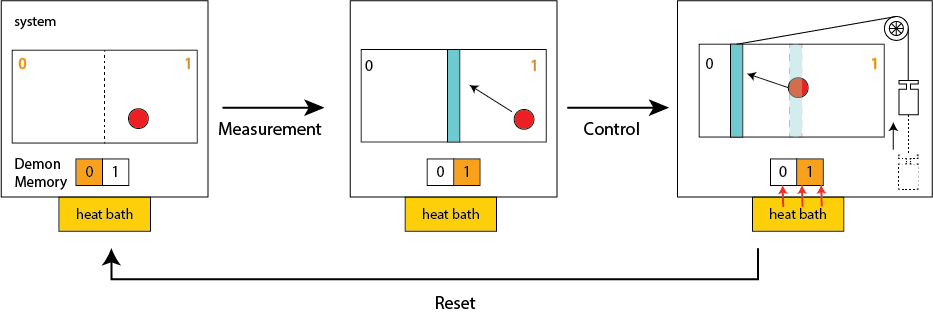}
\caption{Three stages of a single-particle Szilard engine: Initially,
    the particle moves freely within the box such that the particle memory can be either ``0'' or ``1'' and the demon memory is at its default state ``0''. (Step 1) Measurement: A partition is inserted in the middle of the box, confining the particle to the right or left side. If the particle is on the left (right) side, the demon memory is updated to $0$ ($1$). (Step 2) Control: Determined by the measurement result, a weight is attached to the partition in a way such that when the particle pushes the partition, the weight moves upward, storing gravitational potential energy. In short, heat flows from the heat bath into the system to compensate the work done by the particle in lifting the weight. (Step 3) Reset: The partition and the weight are removed and the demon memory is reset to the default state $0$. The engine is ready to enter another cycle.}
\label{fig:szilard_engine_prototype}
\end{figure*}

Section III delves into the QFP implementation, presenting circuit diagrams and extracting the governing equations of motion. The system behavior can be modelled with Langevin dynamics, and the engine implemented by manipulating a potential landscape with external parameters. Section IV explains the metrics used to evaluate the engine's performance: average net work done per cycle ($\langle W \rangle$), transient work costs of information erasure/creation ($W_{erasure}$/$W_{create}$), and information processing fidelity.

Section V presents detailed simulation results for physically-realistic device parameters. While we first consider an idealized device---i.e., ones with exactly identical Josephson junctions---we analyze the consequences of fabrication variability between the Josephson junctions. We find that fabrication variability can significantly affect the device performance. However, we show that the effects can be mitigated by proper control protocols. We also show that incomplete bit erasure and creation have the benefit of reducing $\langle W \rangle$.

Finally, we conclude with Section VI, finding that the results are close to the predicted fundamental values of an ideal Szilard engine. This, in turn, suggests the device will serve as an experimentally-viable testing ground to probe the fundamental energetic consequences of information processing.  

\section{Single particle Szilard Engine}
\label{sec:SzilardEngine}

Consider a system comprising a box containing a single ideal gas particle (Fig. \ref{fig:szilard_engine_prototype}). The box is immersed in a heat bath. An observation and control subsystem, termed a ``demon'' is capable of determining the particle's position and taking actions accordingly. The particle position $x$ can be encoded with a value of $x < 0$ signifying a ``0'' state and $x > 0$ indicating a ``1'' state. In this way, we interpret the particle's position as storing information---the particle ``memory''. Crucially, since the demon needs the ability to distinguish and act on the particle's two states, it must be capable of storing a bit of information---the demon ``memory''. This demon memory is in a default state ``0'' at the beginning of a Szilard engine cycle, with no correlation between particle and demon memory. The engine then operates via a cycle consisting of three distinct stages: measurement, control, and reset. 

During the measurement stage, a partition is inserted into the middle of the box, trapping the particle to one side or the other. The demon memory is updated according to the position of the particle so that the particle memory and the demon memory correlate to each other after the measurement stage. 

For the control stage, a load is attached to the partition according to the result of the measurement process, so that when the particle pushes the partition the particle does work on it. Since the box is connected to a heat bath, the particle's internal energy is fixed. The work done by the particle is compensated by a heat energy transfer from the bath into the system. The engine, in effect, uses the thermal energy from the thermal environment to do work. By the end of these steps, the correlation between particle state and demon memory has been destroyed. 

In the reset stage, the partition and the weight are removed and the demon memory is reset to the default state 0. After that, the system enters a new cycle. 

In the control stage, one bit of information is created in the particle memory. In the reset stage, one bit of information is erased in the demon memory. In this ``single particle'' Szilard engine, the work costs for the measurement, control, and reset process are bound by $0$, $-k_BT\ln2$, and $k_BT\ln2$, respectively \cite{bennett1982thermodynamics, PhysRevE.52.3495, PhysRevA.61.062314, PhysRevLett.102.250602}. It is possible to generalize the single-molecule Szilard engine by varying the parameters involved, resulting in trade-offs between the work costs of the three stages \cite{PhysRevLett.116.190601, ray2020variations}. However, the sum of the work costs is nonnegative since the engine cannot violate the second law. 


\section{Inductively Coupled Quantum Flux Parametrons }
\label{sec:CoupledQFP}

While the engine's basic operating principles are as described, practical operation and observation require a device instantiated on a physical substrate. We propose a substrate based on \emph{quantum flux parametrons} (QFPs). This class of superconducting circuit is well understood, developed over decades of effort in fabrication, control, and usage. Additionally, operating at cryogenic temperatures means that extremely detailed and accurate thermodynamic (calorimetric) measurements can be made---necessary to resolve the $k_B T$-scale energy fluxes on which the engine operates---as irrelevant thermal pathways are frozen out.

Additionally, this implementation does not rely on feedback controls, which are usually implemented by optical-trap wells \cite{PhysRevLett.129.130601, 10.1063/1.5055580, Bustamante2021-ve}, one of the more common alternative platforms for investigating information thermodynamics. It turns out that these feedback loops become a roadblock when trying to measure the minimal costs associated with observing and controlling a nanoscale-scale system. The dominant information costs quickly become those associated with the feedback loop, rather than the operation of the engine under study.

\subsection{Coupled QFP Circuit}

QFP circuits have been used as candidate devices for classical information processing, as well as for measuring fundamental thermodynamic costs \cite{PhysRevResearch.3.033115, PhysRevResearch.3.023164, PhysRevResearch.2.013249, ray2022gigahertz, tanaka201218, 10.1063/1.3585849, 10.1063/1.5080753, PhysRevApplied.4.034007, 10.1063/1.4790276}. One use assigns one of the device's dynamical degrees of freedom $\p$ as an information bearing coordinate as it has two essentially degenerate energy minima under certain control parameter settings. This provides storage for one bit of information. We can thus assign $\p$ to the particle memory---providing a physical implementation of the ``particle'' being on the left ($0$) or right ($1$) of Szilard's box. We choose a second QFP device to serve as the demon memory---also choosing its $\p$ coordinate to store the demon memory information and again naming its stable states as $0$ and $1$.

To allow correlations to build between the two devices, they are coupled with a tuneable inductance, denoted $M_{12}$. Thus, the entire composite system stores the joint particle-demon state in the the two-dimensional space $(\p[1], \p[2])$. (The subscript $i = \{1,2\}$ indexes the first and second QFP device, respectively.) Choosing the same type of device and coordinate to store the information puts them on an equal footing with the work extraction component. This choice enables easily tracking how measurement and control play off of one another with a ``no free lunch'' type trade-off. 

\begin{figure}[ht!]
\includegraphics[scale=0.45]{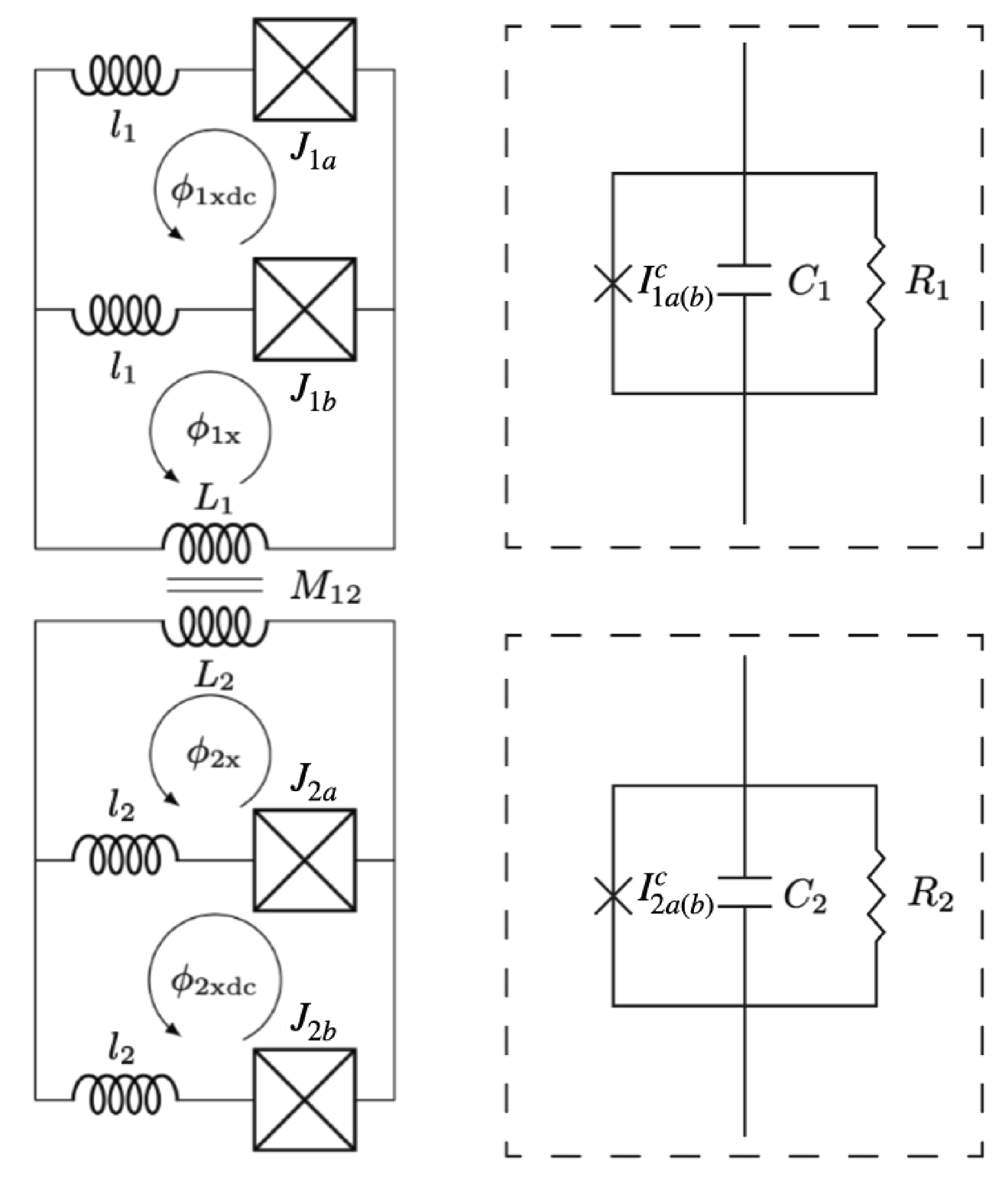}
\caption{Coupled QFP circuit adapted from Fig. 2 of Ref.
    \cite{arXiv:2307.01926v3}. Square boxes $J_{1a}$, $J_{1b}$, $J_{2a}$, and $J_{2b}$ represent  Josephson junctions. Their critical currents are $I^c_{1a}$, $I^c_{1b}$, $I^c_{2a}$, and $I^c_{2b}$, respectively. The diagrams within the dashed boxes represent Josephson junctions modeled by the RCSJ model. $J_{1a}$ and $J_{1b}$ have the same capacitance $C_1$ and resistance $R_1$. Similarly, the capacitance and resistance of $J_{2a}$ and $J_{2b}$ are $C_2$ and $R_2$.}
\label{fig:cfq_circuit_diagram}
\end{figure}

\begin{table} [h]
\begin{subtable}{(a) Fabrication parameters}
\centering
\begin{ruledtabular}
\begin{tabular}{l l l}
\textrm{Symbol}&
\textrm{Physical meaning}&
\textrm{Value}\\
\colrule
 $R_1, R_2$ & Resistance of the JJs & $371 \ \Omega$ \\
 $C_1, C_2$ & Capacitance of the JJs & 4 nF \\
 $L_1, L_2$ & Inductance of two circuits & 0.35 nH \\
 $\ell_1, \ell_2$ & Inductance in series with the JJs & 25 pH \\
 $I^c_{1a}, I^c_{1b}$ & Critical currents of $J_{1a}$ and $J_{1b}$ & $1 ~\mu A$\\
 $I^c_{2a}, I^c_{2b}$ & Critical currents of $J_{2a}$ and $J_{2b}$ & $1 ~\mu A$\\
\end{tabular}
\end{ruledtabular}
\end{subtable}

\begin{subtable}{(b) Calculated parameters}
\centering
\begin{ruledtabular}
\begin{tabular}{l l l}
\textrm{Symbol}&
\textrm{Formula}&
\textrm{Value}\\
\colrule
 $I_{+1}, I_{+2}$ & $I^c_{1a} + I^c_{1b}$, $I^c_{2a} + I^c_{2b}$ & $2~\mu$A\\
 $I_{-1}, I_{-2}$ & $I^c_{1a} - I^c_{1b}, I^c_{2a} - I^c_{2b}$ & $0~\mu$A\\
 $\beta_{1(2)}$ &  $2\pi L_{1(2)} I_{+1(2)} / \Phi_0$ & 2.12\\
 $\delta \beta_{1(2)}$ &  $2\pi L_{1(2)} I_{-1(2)} / \Phi_0$ & 0 \\
 $\gamma_{1(2)}$ & $L_{1(2)}/2\ell_{1(2)}$  & 10 
\end{tabular}
\end{ruledtabular}
\end{subtable}

\begin{subtable}{(c) External parameters}
\centering
\begin{ruledtabular}
\begin{tabular}{l l}
\textrm{Symbol}&
\textrm{meaning}\\
\colrule
 $\px[1], \px[2]$ & Dimensionless flux threading through \\
 & the circuit loop containing $J_a$ and $L_1$ \\ & ($J_c$ and $L_2$) \\
 $\pxdc[1], \pxdc[2]$ & Dimensionless flux threading through \\
 & the circuit loop containing $J_a$ and $J_b$ \\ & ($J_c$ and $J_d$)\\
 $m_{12}$ & Coupling between $L_1$ and $L_2$\\
\end{tabular}
\end{ruledtabular}  
\end{subtable}
\bigskip
\caption{Tables listing the fabrication parameters, calculated
    parameters, and external parameters} 
\label{table:circuit_parameters}
\end{table}

Figure \ref{fig:cfq_circuit_diagram} displays a circuit diagram for the joint device. $J_{1a}$ and $J_{1b}$ ($J_{2a}$ and $J_{2b}$) are the Josephson junctions (JJs) of QFP 1 (2). Similarly, the inductance of the outer loop is $L_{1(2)}$. Both branches inside the loop containing the two JJs have an inductance $l_{1(2)} \ll L_{1(2)}$.

Each JJ is modelled as a \emph{resistive capacitive shunted junction} (RCSJ) \cite{1968ApPhL..12..277S, 10.1063/1.1656743}. With this, $J_{1a}$ and $J_{1b}$ in QFP 1 have an inherent shunt resistance ($R_1$) and a capacitance ($C_1$), and they are connected in parallel with the junctions. The critical currents of $J_{1a}$ and $J_{1b}$ are $I^c_{1a}$ and $I^c_{1b}$. $R_2$, $C_2$, $I^c_{2a}$, and $I^c_{2b}$ are defined similarly for QFP 2. We consider these parameters to be fixed since they are determined at circuit fabrication.

Based on these fabrication parameters, we define several calculated parameters that appear in the effective potential energy of the coupled QFP system. $I_{+1(2)}$ is the sum of critical currents of $J_{1a}$ and $J_{1b}$ ($J_{2a}$ and $J_{2b}$). $I_{-1(2)}$ is the difference of the critical currents of $J_{1a}$ and $J_{1b}$ ($J_{2a}$ and $J_{2b}$). $\beta_1$ and $\beta_2$ are a dimensionless scaling of the maximum flux in the outer loop. $\delta \beta_{1}$ and $\delta \beta_{2}$ yield asymmetric terms in the potential, stemming from manufacturing variance in JJ critical currents. $\gamma_1$ and $\gamma_2$ are the ratio of the value of the large and small inductances. 

Manipulating (i) the external fluxes through the loop and (ii) the coupling between the two QFPs controls the dynamics of the $\p$ and $\pdc$ degrees of freedom. External fluxes $\phi_{1(2)x}$ and $\phi_{1(2)xdc}$ are external magnetic fluxes threading the large inductance loop and the small inductance loop of QFP 1 (2), respectively. $\varphi_{1(2)x}$ and $\varphi_{1(2)xdc}$ are the normalized flux variables defined in the following relations:
\begin{equation*}
 \varphi_{1(2)x} = \frac{2\pi}{\Phi_0} \phi_{1(2)x} \text{ and } \varphi_{1(2)xdc} = \frac{2\pi}{\Phi_0} \phi_{1(2)xdc}, 
\end{equation*}
where $\Phi_0$ is the flux quantum. $M_{12}$ is the coupling between inductors $L_1$ and $L_2$. It is assumed to be tuneable, which can be implemented in practice by using a third SQUID device as the coupler \cite{PhysRevApplied.13.034037, vanDenBrink_2005, PhysRevLett.98.177001, PhysRevB.80.052506}. In the following, $m_{12} = M_{12} / \sqrt{L_1 L_2}$ is the normalized mutual inductance of the circuit.

These external parameters provide the sole means of dynamically modifying the equations of motion (EoM), since the circuit's potential landscape can be directly manipulated with them \cite{ray2022gigahertz, PhysRevResearch.3.023164}. The fabrication parameters, calculated parameters, and external parameters are summarized in Table \ref{table:circuit_parameters}. 

\begin{figure*}
\includegraphics[width=\textwidth]{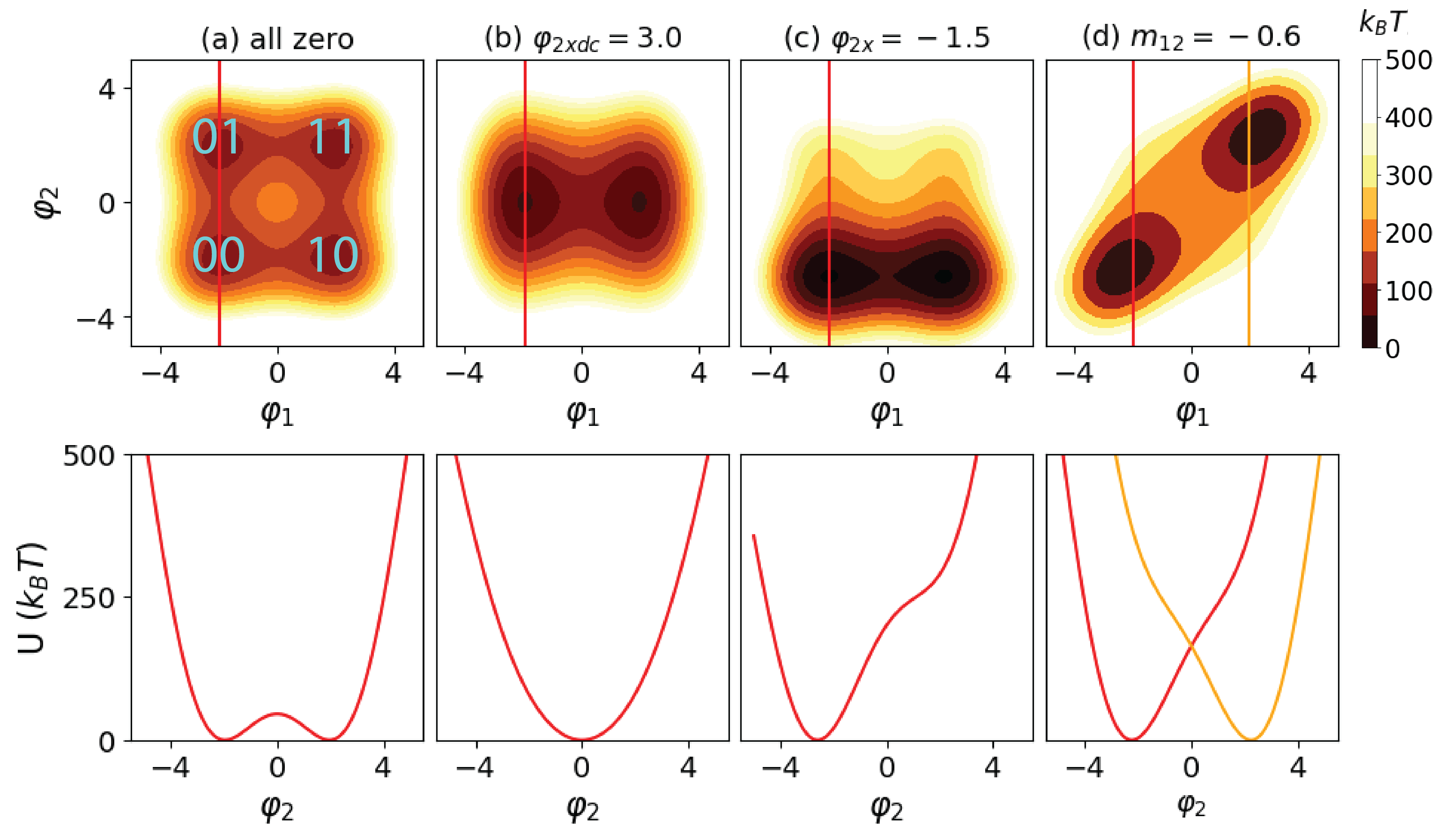}
\caption{Potential landscape in the space of $\p[1]$ and $\p[2]$. The
    color bar shows the energy scale of the contour plots in the unit of $k_BT$. The line graphs below show the energy profile along the red/orange line in the contour plots. (a) When all the external parameters ($\px[1], \px[2], \pxdc[1], \pxdc[2]$, and $m_{12}$) are zero, the energy landscape is a $4$-well potential. The $2$-digit numbers in the contour plot---(00, 01, 10, 11)---label the four wells. (b) $\p[2xdc]$ controls the barrier heights between the top and bottom potential wells. (c) $\p[2x]$ tilts the potential vertically. (d) $m_{12}$ conditionally tilts the potential landscape based on particle location in the left or right half-plane.}
\label{fig:parameter_functions}
\end{figure*}

\subsection{Coupled QFP's Equations of Motion and Potential Landscape}

The system behavior can be modelled by underdamped Langevin dynamics \cite{PhysRevResearch.3.033115, PhysRevResearch.2.013249, variable_beta, PhysRevB.35.4682}, while the engine itself is implemented by manipulating the device's potential landscape with relevant circuit parameters. Reference \cite{arXiv:2307.01926v3} derives the circuit's EoM. After scaling the EoM to be dimensionless (see App. \ref{appendix:EOMs}), we have the following Langevin equation:
\begin{align}
\label{eq:non_dimensionalized_EOMs}
dv'_i = - \lambda_i v'_i dt' - \theta_i \partial_{x'_i} U' + \eta_i \ r_i(t) \sqrt{2dt'}, 
\end{align}
for $i = 1, 2, 3, 4$. The prime notation on $x'_i$, $v'_i$, $U'$, and $t'$ denotes them as dimensionless scalings of the position, velocity, potential, and time. Of primary importance are the effective degrees of freedom in the circuit ($x'$, above) which are related to the junction phases. Let the junction phases of the JJ $J_{1a}$, $J_{1b}$, $J_{2a}$, and $J_{2b}$ be $\delta_{1a}$, $\delta_{1b}$, $\delta_{2a}$, and $\delta_{2b}$. The dynamical coordinates can be expressed in terms of the JJ phases as:
\begin{align*}
 x'_{1(2)} = \varphi_{1(2)} &= \frac{\delta_{1(2)a} + \delta_{1(2)b}}{2} \\
 x'_{3(4)} = \varphi_{1(2)dc} &= \delta_{1(2)a} - \delta_{1(2)b}
 ~. 
\end{align*}

The Langevin coefficient $\lambda_i, \theta_i$, and $\eta_i$ are defined as components of the following vectors:
\begin{align*}
\vec{\lambda} &= (\frac{2}{R_1} \sqrt{\frac{L_1}{C_1}}, \frac{2}{R_2} \sqrt{\frac{L_2}{C_2}}, \frac{2}{R_1} \sqrt{\frac{L_1}{C_1}}, \frac{2}{R_2} \sqrt{\frac{L_2}{C_2}}) ,\\
\vec{\theta}  &= (1, \frac{C_2}{C_1}, 4, 4\frac{C_2}{C_1}), ~\text{and}\\
\vec{\eta}    &= \sqrt{\frac{k_BT }{U_0}} (\sqrt{\lambda_1}, \sqrt{\frac{C_2\lambda_2}{C_1}}, 2\sqrt{\lambda_3}, 2\sqrt{\frac{C_2\lambda_4}{C_1}})
  ~.
  \end{align*}

$\vec{\lambda}$ is the thermal coupling coefficient, reflecting the strength of the damping force the particles experience from the thermal bath. $\vec{\theta}$ is the strength of the potential relative to the system dynamic. $\vec{\eta}$ is related to the strength of the thermal white noise $r_i(t)$ and $U_0 = \Phi_0^2/(4 \pi^2 L_1) = \Phi_0^2/(4 \pi^2 L_2)$.


\subsubsection{Dimensionless Potential Energy}

Reference \cite{arXiv:2307.01926v3} gives the full dimensional equations of the potential of the coupled QFPs. Appendix \ref{appendix:potential_equal} derives the following nondimensional potential $U'$ in Eq. (\ref{eq:non_dimensionalized_EOMs}). $U'$  can be expressed as:
\begin{align} \label{eq:potential of coupled flux qubits}
U' = U_1 + U_2 + U_{12}
\end{align}
with:
\begin{align*}
    U_i &= \frac{\xi}{2}(\p[i] - \px[i])^2 + \frac{\gamma_i}{2}(\pdc[i]-\pxdc[i])^2 \\
    & +\beta_i \cos\p[i] \cos\frac{\pdc[i]}{2} 
     + \delta \beta_i \sin\p[i] \sin\frac{\pdc[i]}{2} \\
    U_{12} &= m_{12} \xi(\p[1] - \px[1])(\p[2] - \px[2]),\\
\end{align*}
where $\xi \equiv \frac{1}{1 - m_{12}^2}$.

Notably, $\gamma_1$ and $\gamma_2$ are typically larger than $\xi$. For the device in Sec. \ref{sec:Sims}, the value of $m_{12}$ varies between 0 and 0.6 throughout the protocol and the ratio $\frac{\gamma_i}{\xi} \approx 6$ falls within the range of 6 to 10. If the value of $\pdc[1](\pdc[2])$ deviates from the external flux $\pxdc[1](\pxdc[2])$, the system pays a large energy penalty when compared to $\p[1](\p[2])$ deviating from $\px[1](\px[2])$. Due to this penalty, we can assume that $\pdc[1]$ and $\pdc[2]$ closely match $\pxdc[1]$ and $\pxdc[2]$, throughout the operation. This assumption reduces the dominant dynamics from a $4$-dimensional to the $2$-dimensional subspace defined by $\p[1]$ and $\p[2]$ only. This simplifies the design space of the protocol significantly. That being said, while the protocol is designed with only the $(\p[1],\p[2])$ subspace in mind, we simulate using the the full $8$-dimensional phase-space dynamics.

\subsubsection{Szilard Engine Control Protocol}

When all the external parameters are zero, the relevant 2D potential energy in the $\p[1]$-$\p[2]$ space is a $4$-well potential (Fig. \ref{fig:parameter_functions}a). The energy barriers separating the potential wells are substantially larger than the thermal energy $k_BT$ and so prevent undesirable jumps between the wells. In this way, the minima provide metastable regions that serve as reliable memory states with escape rates of about $10^{-43}\text{Hz}$ at $T=0.5$ K. (See App. \ref{appendix:escape_rate}.) Recall that we use $\p[1]$ to indicate the state of the particle: for $\p[1] < 0 \ ( \p[1] > 0)$, the particle is in state $0$ ($1$). Meanwhile, $\p[2]$ acts as the demon memory: For $\p[2] < 0 \ ( \p[2] > 0)$, the demon memory is $0$ ($1$). The effect that each control parameter has on the potential is shown in Fig. \ref{fig:parameter_functions}: $\varphi_{1(2)xdc}$ controls the barrier height between the potential wells (Fig. \ref{fig:parameter_functions}b), $\varphi_{1(2)x}$ tilts the potential landscape (Fig. \ref{fig:parameter_functions}c.) $m_{12}$ causes a ``conditional tilt'' that acts differently in the left semiplane than right one, tilting each in an opposite direction (Fig. \ref{fig:parameter_functions}d). These three operations can establish a protocol that replicates the operation of the Szilard engine within the $(\p[1],\p[2])$ subspace of the full phase space of the coupled QFP system.

\begin{figure*}
\includegraphics[scale=0.83]{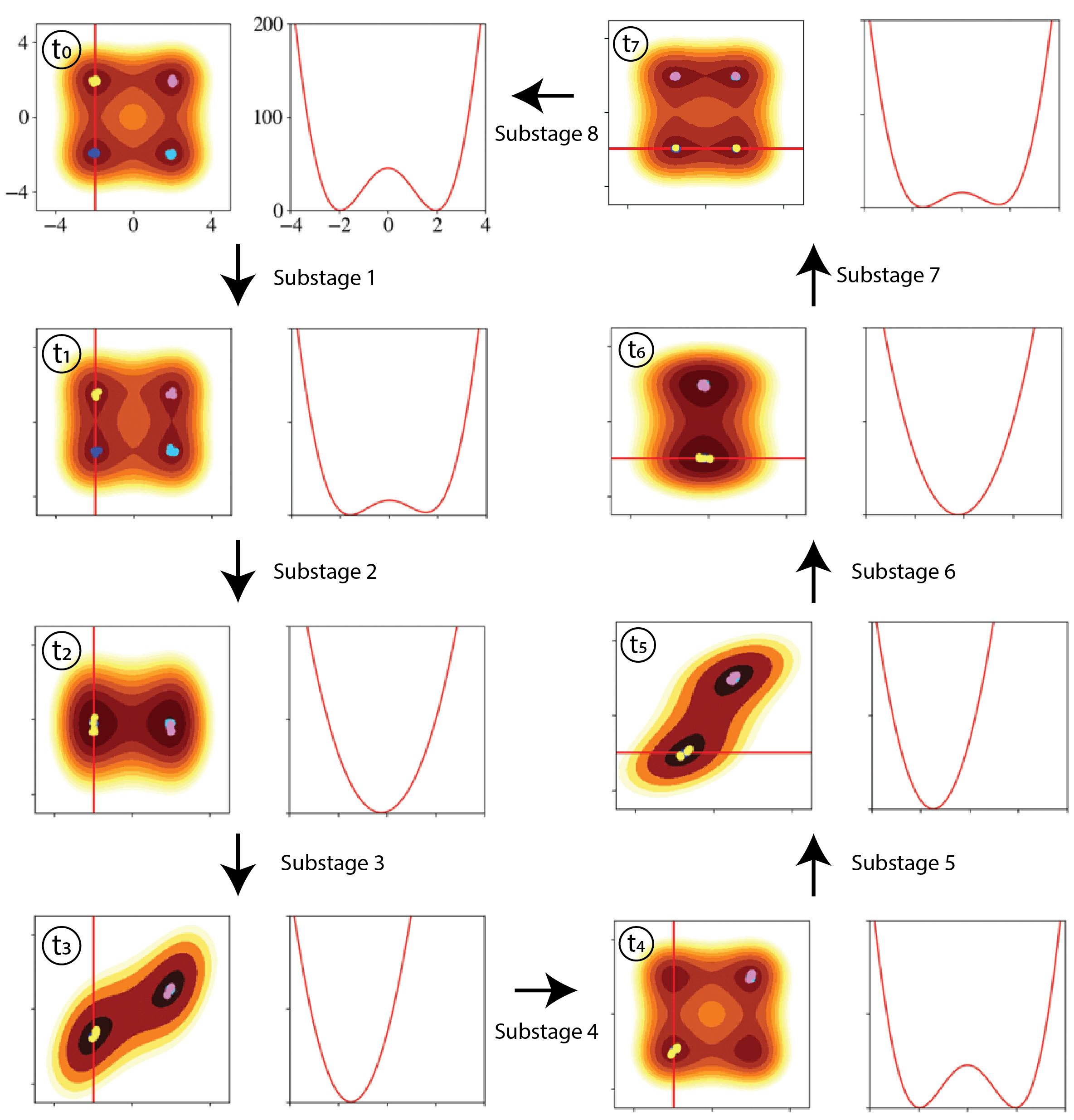}
\caption{Evolution of the potential---snapshots at different cycle 
    substages: The $x$-axis and $y$-axis of the contour plots are $\p[1]$ and $\p[2]$, respectively. The color dots show the distribution of the particles at different times. The dark blue, yellow, light blue, and pink particles are the particles with initial state inside region $00$, $01$, $10$, and $11$, respectively. The line graphs show the potential along the red cutlines in the contour graphs. The $x$-axis is $\p[1]$ or $\p[2]$ depending on the direction of the cutlines, and the $y$-axis is the potential energy in units of $k_BT$. The red cutlines pass through the local minima of the potential. (\textcolor{blue}{\href{https://drive.google.com/file/d/117_rGZXyq3inFWFaBFVCs6DgFsLUoi0k/view?usp=drive_link}{Animation available online}})}.
\label{fig:substage_potential_and_distribution}
\end{figure*}

Figure \ref{fig:substage_potential_and_distribution} guides one through each step. The particle colors index the initial wells in which they start: Dark blue for $00$, yellow for $01$, light blue for $10$, and pink for $11$. Note that the demon memory ($\p[2]$) is not set to the default state ``0'', rather our start state is the equilibrium distribution associated with the initial potential energy profile. This choice is just as valid as a deterministic demon start state for the memory, since there is no correlation between the particle state and the memory. That is, a particle on the left is just as likely to be in well $00$ as in well $01$.

During Substage 1 and Substage 2 ($t_0$ to $t_2$)  the potential barrier between the top and bottom wells is lowered. This allows the initially separated dark blue and yellow (light blue and pink) particles to intermix. Substage 3 ($t_2$ to $t_3$) is a conditional tilt of the potential along the $\p[2]$ direction to correlate the state memory and the demon memory. In Substage 4 ($t_3$ to $t_4$), the potential returns to its original $4$-well potential. There was no correlation between the state memory and the demon's memory at $t_0$, but by $t_4$ they have become perfectly correlated. Therefore, Substages 1-4 represent how the protocol implements measurement. During this stage, one bit of information is erased as each particle's accessible state space is effectively halved. 

Substage 5 ($t_4$ to $t_5$) is a conditional tilt of the potential along the $\p[2]$ direction and Substage 6 ($t_5$ to $t_6$) returns us to a quadratic potential along $\p[2]$ direction. These two substages allow the intermixed dark blue and yellow (light blue and pink) particles to relocate to the center between regions $00$ and $10$ (regions $01$ and $11$). Substage 7 and Substage 8 ($t_6$ to $t_8$) reraise the barrier between the left and right wells, returning to the symmetric $4$-well potential. Substages 4-8 represent the control process of the protocol: the correlation established in measurement process is harvested and one bit of information is created---each particle's accessible state space is effectively doubled. The work extracted between Substage 5 and Substage 8 represents the work gain during the protocol's control stage.

No reset process is required in this protocol since, at the end of the control stage, the system is already primed for another cycle. This is simply due to our choice of initial state, as opposed to anything fundamental. The parameter values used to generate Fig. \ref{fig:substage_potential_and_distribution} and the simulations in the next section are summarized in Fig. \ref{fig:protocol_values} of App. \ref{appendix:simulation_detail}.

\begin{figure*}
\includegraphics[scale=0.57]{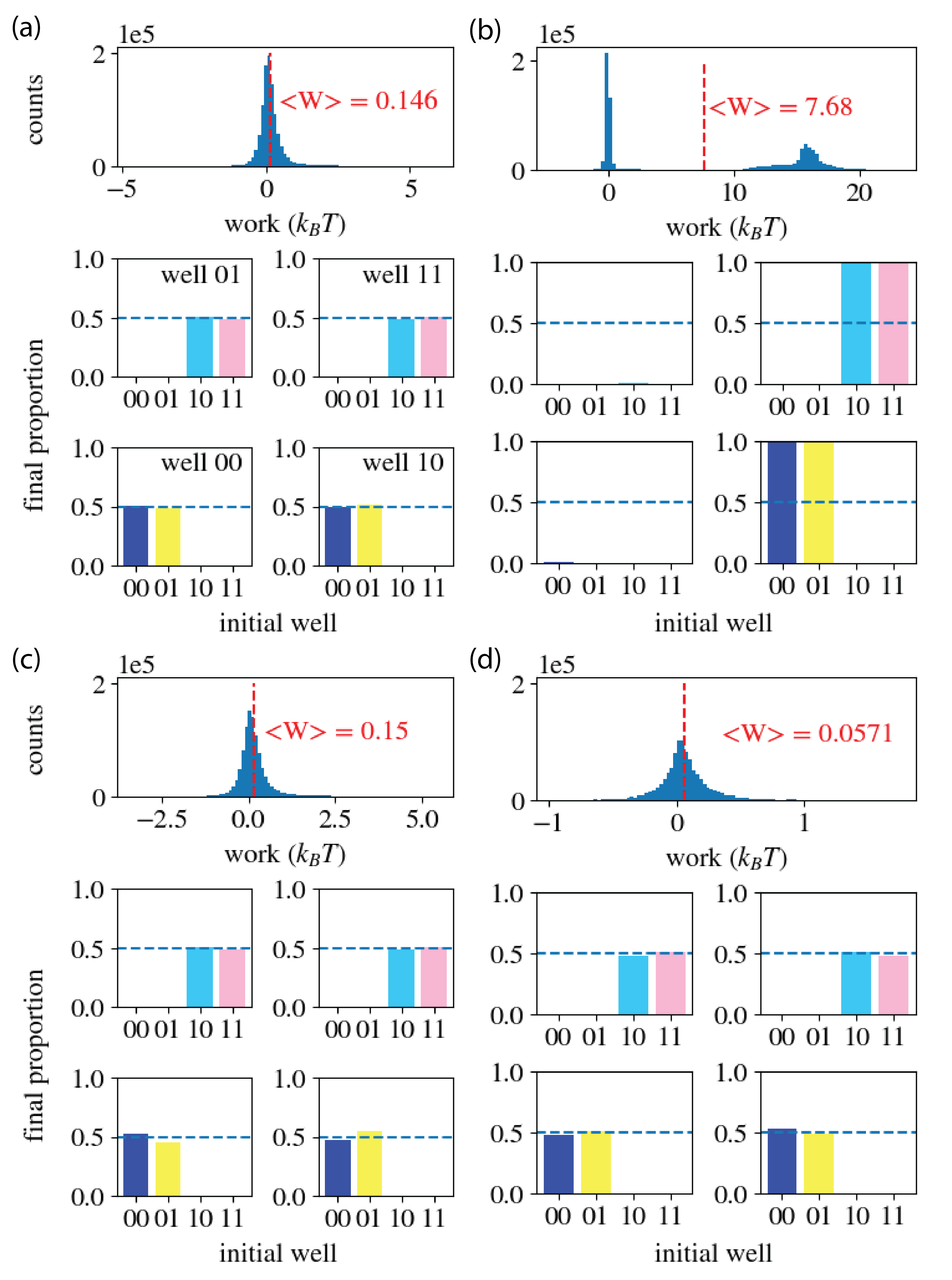}
\caption{The final work distributions and the final particle
    proportions of $N = 10^6$ particles in each of the four wells for the four cases: (a) symmetric device, (b) asymmetric device with no tilting, (c) asymmetric device with tilting; and (d) symmetric device without substage 4 in Fig. \ref{fig:substage_potential_and_distribution}. Initial well identifies in which well the particle starts. The red dashed lines indicate the average total work, and the blue dashed lines in the particle proportion plots indicate the value of $0.5$.}
\label{fig:work_distribution_and_fidelity}
\end{figure*}

\begin{figure*}
\includegraphics[scale=0.55]{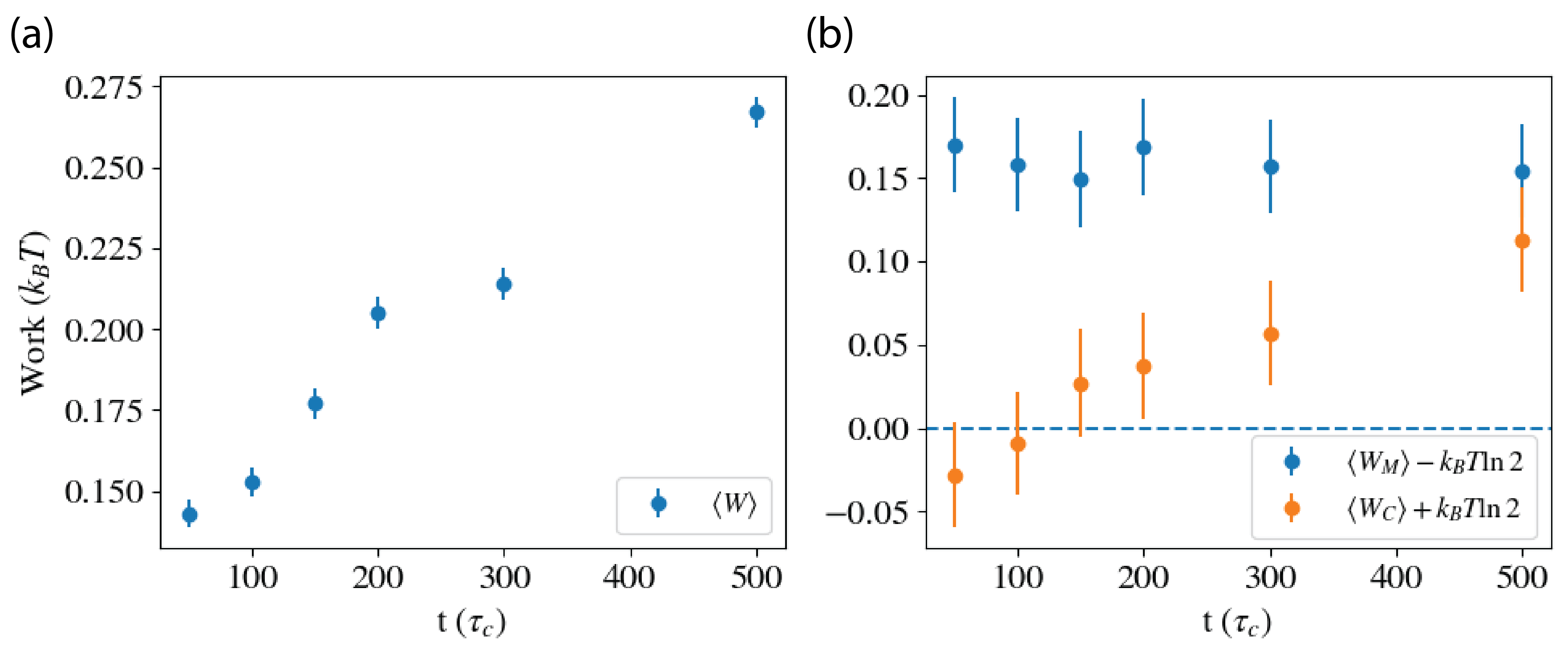}
\caption{(a) The average total work $\langle W \rangle$ and (b) work
    deviations for measurement and control (relative to their respective bounds $\pm k_BT\ln2$) as the duration of substage 4 and substage 5 change from $50$ to $500$. In the ideal case, the deviations vanish, which level is indicated by the dash line. The total work increases and the deviations from the bounds converge as substage durations increase.}
\label{fig:step_4_and_5__longer_duration}
\end{figure*}
\section{Results and Discussion}
\label{sec:Sims}


The previous section established the device parameters, the system's dynamical equations, and its potential landscape. It also introduced the control protocol for the device to perform as a Szilard engine. The following now presents the simulation results obtained through underdamped Langevin dynamics. To evaluate the engine's performance, we employ two primary figures of merit: the net work cost and the engine's functional fidelity. The average total work cost quantifies the total energy expenditure required to operate the engine through a complete cycle. Fidelity serves as a measure of how accurately the engine performs. Additionally, substage work costs are detailed as a secondary analysis tool.

The proposed QFP Szilard engine has substage work cost bounds that are distinct from the Szilard engine described in Sec. II. This difference arises from the choice of the initial demon memory state. In the original Szilard engine, the memory is initialized to the default 0 logical state. However, in the QFP engine, the demon's memory is set to the equilibrium state of a 4-well potential. This modification transfers the burden of erasure cost from the reset process to the measurement process, because the uncorrelated demon memory is now erased during the measurement process.

Comparing the particle distribution at $t_0$ and $t_4$ in Fig. \ref{fig:substage_potential_and_distribution}, the state space volume decreased by half. By Liouville's theorem, the total phase space volume must be conserved and, thus, the decrease in state space volume of the system must be compensated by the expansion of the state space volume of the heat bath. On the one hand, the change in entropy of the system is $-k_B\ln2$ and the minimum amount of heat the system must dump into the heat bath is $k_BT \ln 2$. On the other hand, after the control substage, the state space volume of system is doubled and so the change in entropy of the system is $k_B\ln2$. The maximum amount of heat the system absorbs from the heat bath is $k_BT \ln2$. These values are consistent with Landauer's principle. Despite the difference in substage work, the net work is still bound by zero in these two types of Szilard engine, since the processes are both cyclic. 

Of course, any finite-time process costs more work than the quasistatic bound, but these two values can be used as reference points to check the efficiency of the control protocol's substages. We now consider cases that explore the role of landscape symmetry and imperfect operation on the thermodynamics.

\subsection{Symmetric Potential}

If we assume idealized JJ fabrication in which each JJ can be paired with a twin that has identical critical current (i.e., $I_{-1(2)} = 0 \mu$A), then $\p[1x]$ and $\p[2x]$ are both $0$ throughout the whole cycle. While the fabrication of two such JJ elements is likely infeasible, we investigate this case first, as a point of reference.

Figure \ref{fig:work_distribution_and_fidelity}(a) shows the net work distribution, which is a single peak distribution with mean work $0.146$ $k_BT$. A genuinely adiabatic/quasistatic operation would yield zero net work, as the process is cyclic. However, the protocol is a finite-time process, so the work cost is positive. The first row of Table \ref{table:work_cost} shows that the work done for the measurement process ($W_M$) and that for the control process ($W_C$) are both within $0.13 ~k_BT$ of their fundamental bounds. The work done for all substages are shown in Table \ref{table:substage_mean_work} in App. \ref{appendix:substage_work_distribution}

Interestingly, we see a difference between the work done for the measurement and control processes relative to their respective bounds. The measurement process deviates from its bound by approximately $0.13 k_BT$, while the control process deviates by about $0.042 k_B T$. This asymmetry arises due to the system's nonequilibrium behavior during the protocol. 

To investigate this effect, we extended the durations of substages 4 and 5 from $t = 50$ to $t = 500$, in units measured in $t_c = \sqrt{LC} \approx 1~\text{ns}$. As is evident from Fig. \ref{fig:step_4_and_5__longer_duration}, increasing the protocol duration leads to more symmetric deviations from the bounds for both processes. This is because the entire process becomes more adiabatic with a longer duration. Note, also, that increasing the durations results in a diminishing return of extractable energy from the measurement stage. This decrease in recovered energy translates to an overall increase in the average total work done due to less work being retrieved during the control process. It is worth emphasizing, as a side point, that the nonequilibrium behavior displayed here is to the benefit of the overall engine: a surprising result given the protocol was designed with the (local) equilibrium distributions in mind.

Figure \ref{fig:work_distribution_and_fidelity}(a) depicts the final proportion of the particles in each of the four wells. In the ideal situation, the dark blue and yellow (light blue and pink) particles should distribute evenly in the well $00$ and $10$ ($01$ and $11$), as seen in Fig. \ref{fig:work_distribution_and_fidelity}(a). Thus, the circuit supports a low-cost, high-fidelity implementation of the Szilard engine.

\subsection{Compensated Asymmetry}
\label{sec:CompAsymm}

The previous section provided a performance baseline by assuming an ideal symmetric device. However, due to fabrication variations \cite{7792220, Zeng_2015}, there is always a difference in JJ critical currents. Indeed, the difference can be up to 8\% \cite{asymmetryJJ6967749}. This section assumes instead an inexact fabrication by setting the difference in the JJ critical currents to $I_{-1(2)} = 0.16 \mu A$ ($\delta \beta \approx 0.170$)---a conservative 8\% difference in the critical currents of the junctions. Figure \ref{fig:work_distribution_and_fidelity}(b) shows the final work distribution and final proportion of particles inside the four wells. 

We see that the $8\%$ fabrication variability completely destroys the intended protocol. The final distribution is nowhere close to correct. Nearly all particles relocated to the potential wells on the right side. Additionally, the average work is about $60$ times larger than the symmetric case and the work distribution has two very distinct peaks. The engine failed to do its intended task, and it has done so at great expense. Something has gone very wrong. 

\begin{figure}[h]
\includegraphics[scale=0.25]{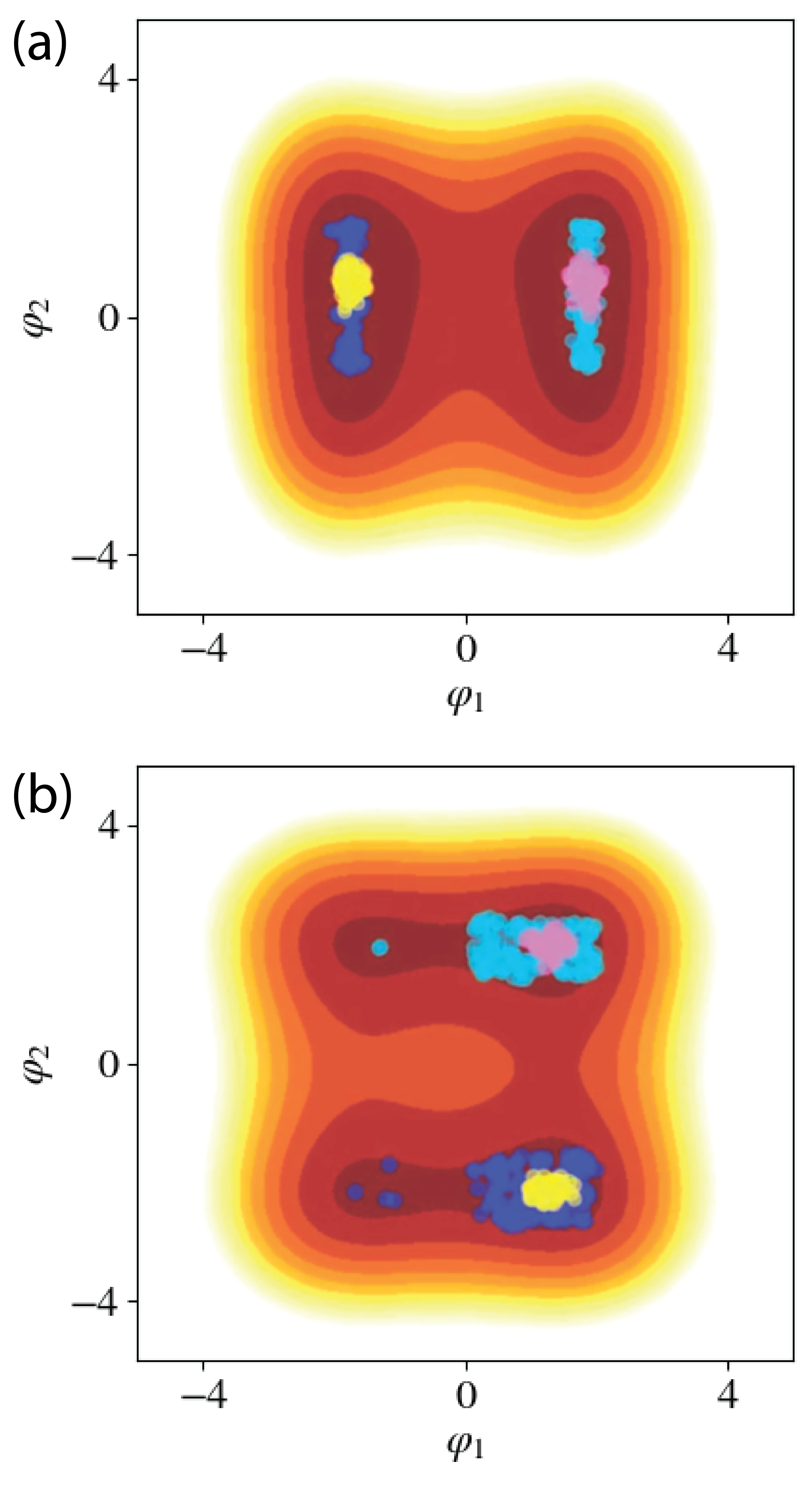}
\caption{Particle distribution at the end of (a) Substage 2 and (b)
    Substage 8 for the case of asymmetric potential without compensation. The dark blue and light blue particles remain far away from local equilibrium throughout the process due to the asymmetric potential gradient introduced by the asymmetric Josephson junctions. These particles contribute to the high-work trajectories observed in the final work distribution in Fig. \ref{fig:work_distribution_and_fidelity}(b). (\textcolor{blue}{\href{https://drive.google.com/file/d/1S2XKeGgEdYz106vrhUaqLfIrjkc43j-w/view?usp=drive_link}{Anmiation available online}})
    }
\label{fig:final_distribution_asymmetric}
\end{figure}

It is fairly straightforward to see the problem. The asymmetry in the critical current leads to an asymmetric potential energy term, causing an inherent tilt that favors transitions upwards and to the right. This tilt results in uneven behavior of particles across the four wells and an uneven distribution at the end of a cycle. When the potential well is lowered, the additional tilt causes the dark blue and light blue particles to gain extra kinetic energy as they are transitioning in the favored direction. Figure \ref{fig:final_distribution_asymmetric}(a) shows that these particles have been driven far away from local equilibrium. Conversely, the pink and yellow particles oppose the potential gradient during the lowering phase, gaining negligible kinetic energy.

As a result, the pink and yellow particle distributions remain close to local equilibrium throughout the process. Meanwhile, the blue and dark blue particles are still far away from local equilibrium at the end of the protocol (Fig. \ref{fig:final_distribution_asymmetric}(b)) because the dynamics are very underdamped. This stark contrast in particle behavior leads to distinct work outcomes. While pink and yellow particles incur minimal work cost, the dark blue and light blue particles experience substantial work expenditure. Consequently, the net work distribution displays a two-peak pattern.

\begin{table*}
\centering
\begin{tabular}{ | wc{4.5cm} | wc{2.5cm}| wc{2.5cm} | wc{2.5cm} |} 
\hline
Case & $\langle W \rangle$ ($k_BT$) & $\langle W_M \rangle$ ($k_BT$) & $\langle W_C \rangle$ ($k_BT$) \\
\hline
Symmetric & $0.146 \pm 0.001$ & 0.857 ± 0.003 & -0.716 ± 0.003\\
\hline
Asymmetric, uncompensated & 10.4 ± 0.2 &  10.7 ± 0.3 & -0.306 ± 0.460\\
\hline
Asymmetric, compensated & 0.150 ± 0.001  &  0.830 ± 0.003 & -0.679 ± 0.003\\
\hline
Symmetric, no substage 4  & 0.057 ± ± 0.001 & 0.373 ± 0.002 & -0.316 ± 0.002 \\
\hline
\end{tabular}
\caption{Total work $\langle W \rangle$, work for measurement
    $\langle W_M \rangle$, and work for control $\langle W_C \rangle$ for (i) symmetric potential, (ii) asymmetric potential without compensation, (iii) asymmetric potential with compensation, and (iv) symmetric potential without returning to the $4$-well potential after the measurement process. For the last case, the work for measurement and the work for control are defined as the work involved from $t_0$ to $t_3$ and the work involved in the rest of the protocol after substage 4 is removed, respectively.}
\label{table:work_cost}
\end{table*} 

To mitigate this problem, $\p[1x]$ and $\p[2x]$ can be set to a nonzero value (see Fig. \ref{fig:protocol_values}) to tilt the potential, counteracting the effect of the asymmetric term. The values were chosen through a simple parameter sweep $\p[1x]$ and $\p[2x]$ to maximize fidelity. Figure \ref{fig:work_distribution_and_fidelity}(c) shows the final work distribution and final proportion of particles inside the four wells using the optimal correction values of $\p[1x]$ and $\p[2x]$. The average total work and the final proportion of particles for the compensated case are very close to the ideal symmetric case. Thus, we see that $\p[1x]$ and $\p[2x]$ can be used effectively as calibration parameters to account for fabrication variability in the JJ critical currents.

\subsection{Incomplete Bit Erasure and Creation}

Recall that the purpose of the protocol's ``measurement'' process---the first 4 substages of the protocol---is to establish a correlation between the particle state and demon memory. However, after substage 3, the particle state and the demon memory have already become perfectly correlated. That is, knowing the value of one, the other is known with certainty. Thus, substage 4 does not play an important role in establishing this correlation. The substage was actually introduced only to ensure the potential at $t_0$ and the potential after the measurement process are identical. This allowed a fair comparison with the ``standard candle'' of information processing: the Landauer bound on bit erasure and creation, which assumes a cyclic process.

That noted, such an additional step is likely an inefficient use of available resources and, moreover, could become very costly for complex computations. After removing substage 4, one no longer expects the ``measure'' and ``control'' steps to be bounded by the minimal costs of bit erasure and creation, but by changes in the nonequilibrium free energy \cite{parrondo2015thermodynamics}. Regardless, since we are simply removing unnecessary steps, the shortened protocol should achieve a lower net work cost.

Figure \ref{fig:work_distribution_and_fidelity}(d) shows the final work distribution and final proportions of the particles in the four wells. The final proportions of the particles in this case are nearly the same as that in Fig. \ref{fig:work_distribution_and_fidelity}(a). The last row of Table \ref{table:work_cost} show the average total work, work for measurement, and work for control having removed substage 4. This results in incomplete bit erasure and bit creation and that, in turn, means less work is needed to measure and less work recovered during control. Comparing these values with the first case we see a lower average total work, but the net work when all substages are considered is still bounded by $0$.

\section{Conclusion}
\label{sec:Conclusion}

We presented a physically-realistic control protocol for the physical realization of a Szilard engine utilizing inductively coupled QFP circuits. Simulation results demonstrated that this $2$-bit logical unit, based on the quantum flux parametron, is capable of performing highly-efficient information processing with work costs close to the fundamental bounds and with very high fidelity. Helpfully, the physical substrate is a well-understood system that requires no continuous monitoring to be properly controlled. As such, it is well suited for careful measurements of the minimal costs associated with information processing. Furthermore, we tested the impact of asymmetric Josephson junction fabrication on the device's performance through simulation and proposed a solution to mitigate detrimental effects.

The resulting circuit family comes with verifiable protocols to support future experimental investigations into the information thermodynamics and trade-offs associated with Szilard engines, as well as related with devices supporting net positive thermodynamic benefits, such as the \emph{information ratchets} \cite{Boyd15a}. Successful experimental validation will support the development of a new generation of highly energy-efficient information-processing architectures.

\begin{acknowledgments}
The authors thank the Telluride Science Research Center for its hospitality during visits and the participants of the Information Engines workshop there for their valuable feedback. J.P.C. acknowledges the kind hospitality of the Santa Fe Institute and California Institute of Technology. This material is based on work supported by, or in part by, the U.S. Army Research Laboratory and U.S. Army Research Office under Grant No. W911NF-21-1-0048.
\end{acknowledgments}

\appendix

\section{Underdamped Langevin Equations}
\label{appendix:EOMs}

Underdamped Langevin dynamics can often be used to model the behavior of the flux degrees of freedom when JJs are modeled by the RSCJ model. 

The dimensional Langevin equation is:
\begin{equation}
\label{langevin_equation_1}
m_i dv_i + \nu_i v_i dt = -\partial_{x_i}
U(\vec{x}; \vec{\lambda}_{ext}(t)) dt + r_i(t) \sqrt{2 \nu_i \kappa dt}
  ~.
\end{equation}
Here, $U(\vec{x}; \vec{\lambda}_{ext}(t))$ is the driving potential and $\vec{\lambda}_{ext}(t)$ is the protocol of the external parameters. $\kappa$ is equal to $k_BT$. The symbols $x_i$, $v_i$, $m_i$, $\nu_i$, and $r_i$ represent the i-th components of the $\vec{x}$ (position), $\vec{v}$ (velocity), $\vec{m}$ (mass), $\vec{\nu}$ (damping), and $\vec{r}$ (Gaussian random number), respectively, so:
\begin{align*} 
\vec{x} &= (\phi_1, \phi_2, \phi_{1dc}, \phi_{2dc})\\
\vec{v} &= (\frac{d \phi_1}{dt}, \frac{d \phi_2}{dt}, \frac{d \phi_{1dc}}{dt}, \frac{d \phi_{2dc}}{dt})\\
\vec{m} &= (C_1, C_2, \frac{C_1}{4}, \frac{C_2}{4})\\
\vec{\nu_i} &= (\frac{2}{R_1}, \frac{2}{R_2}, \frac{1}{2R_1}, \frac{1}{2R_2})\\
\vec{r} &= (r_1, r_2, r_3, r_4)
  ~.
\end{align*}
In this, the $r_i(t)$ are the 4 independent memoryless Gaussian random variables with zero mean and unit variance. Note that $r_i(t)$ have the following relation:
\begin{align}\label{etaTimeCorrelations}
\langle r_i(t) r_j(t') \rangle = \delta_{ij}\delta(t-t').
\end{align}
The average $\langle \ldots \rangle$ is taken over the stochastic process' trajectories. 

This appendix shows how to obtain the non-dimensional equation, Eq. (\ref{eq:non_dimensionalized_EOMs}) in the main text, from the above dimensional equation. Rearranging terms of Eq. (\ref{langevin_equation_1}) gives:
\begin{align} \label{eq:langevin_equation_2}
dv_i = -\frac{\nu_i}{m_i} v_i dt - \frac{1}{m_i} \partial_{x_i} U(\vec{x})dt + \frac{r_i(t)}{m_i} \sqrt{2 \nu_i \kappa dt}
\end{align} By introducing the following constants:
\begin{align*} 
x_c &= \frac{\Phi_0}{2\pi}; v_c = \frac{x_c}{t_c}; t_c = \sqrt{LC};\\
m_c &=C; \nu_c = \frac{1}{R}; ~\text{and}\\
\kappa_c &= U_0 = \frac{m_c x_c^2}{t_c^2} = \frac{\Phi_0^2}{4 \pi^2 L},
\end{align*}
the Langevin equation can be nondimensionalized, simplifying the simulation. Factoring out these constants from the corresponding variables gives $x_i'$, $v_i'$, $m_i'$, $\nu_i'$, $U'$, and $t_c$ as nondimensional variables. The vectors above become:
\begin{align*} 
\vec{x} &= x_c \vec{x'} = \frac{\Phi_0}{2\pi} (\p[1], \p[2], \pdc[1], \pdc[2]),\\
\vec{v} &= v_c \vec{v'} = \frac{x_c}{t_c} (\frac{d\p[1]}{dt}, \frac{d\p[2]}{dt}, \frac{d\pdc[1]}{dt}, \frac{d\pdc[2]}{dt}),\\
\vec{m} &= m_c \vec{m'} = C_1 (1, \frac{C_2}{C_1}, \frac{1}{4}, \frac{1}{4}\frac{C_2}{C_1}), ~\text{and}\\
\vec{\nu_i} &= \nu_c \vec{\nu_i'} = \frac{1}{R_1} (2, 2\frac{R_1}{R_2}, \frac{1}{2}, \frac{1}{2} \frac{R_1}{R_2}),
  ~
\end{align*} $U = U_0 U'$  and $\kappa = \kappa_c \kappa'$. With these constants, (\ref{eq:langevin_equation_2}) can be expressed as:
\begin{multline}
    v_c dv' = - \frac{\nu_c \nu'}{m_c m'} v_c v' t_c dt' - \frac{1}{m_cm'} \frac{U_0 }{x_c} \partial_{x'} U' dt' \\ 
    + \frac{1}{m_c m'}r(t') \sqrt{2 \nu_c \nu' \kappa_c \kappa' t_c dt'}.
\end{multline}
Note that subscript $i$ in the expression is dropped to simplify the expression.
Dividing by $v_c$ on both sides and making use of $v_c = x_c/t_c$, the above can be simplified:
\begin{multline}
    dv' = - \frac{\nu_c t_c}{m_c } \frac{v'}{m'} v' dt' - \frac{1}{m'}  \partial_{x'} U' dt' \\
    + \frac{1}{m_c}  \sqrt{ \frac{\nu_c \kappa_c t_c^3} {x_c^2} } \frac{\sqrt{\nu' \kappa'}}{m'} r(t') \sqrt{2dt'}
\end{multline}
This can then be further simplified to: 
\begin{align} \label{langevin_equation_3}
dv' = - \lambda v' dt' - \theta \partial_{x'} U'dt' + \eta r(t') \sqrt{2dt'}
    ~.
\end{align}
Here, $\lambda$ is the thermal coupling coefficient, reflecting the strength of the damping force the particles experience from the thermal bath. $\theta$ is related to the strength of potential to the dynamic of the system. If $\theta$ is zero, it means the system is off from the potential. $\eta$ is related to the strength of the noise.

Returning subscript $i$ gives the dynamics in the form of Langevin equations:
\begin{equation} \label{electrostatic}
dv'_i = - \lambda_i v'_i dt' - \theta_i \partial_{x'_i} U' dt' + \eta_i r_i(t') \sqrt{2dt'}
  ~,
\end{equation}
where $\lambda_i, \theta_i$, and $\eta_i$ are defined as:
\begin{align*}
\lambda_i &=  \frac{\nu_c t_c}{m_c } \frac{\nu_i'}{m_i'}\\
\theta_i &= \frac{1}{m_i'}\\
\eta_i &= \frac{1}{m_c}  \sqrt{ \frac{\nu_c \kappa_c t_c^3} {x_c^2} } \frac{\sqrt{\nu' \kappa'}}{m'} = \sqrt{\frac{\lambda_i \kappa'}{m'}}
  ~.
\end{align*}

\section{Potential Landscape for Strong Coupling}
\label{appendix:potential_equal}

For small $m_{12} = M_{12} / \sqrt{L_1L_2}$ it is common to use linear approximation to approximate the equation of the potential. We use the exact form because the value of $m_{12}$ can go up to 0.6. The derivation of the exact form is given here. 

According to Eq. (18) in Ref. \cite{arXiv:2307.01926v3} the potential can be written in the form:
\begin{align*} 
U = &E_{2+1} \cos\p[1] \cos\frac{\pdc[1]}{2} + E_{2-1} \sin\p[1] \sin\frac{\pdc[1]}{2} \\
&+E_{4+3} \cos\p[2] \cos\frac{\pdc[2]}{2} + E_{4-3} \sin\p[2] \sin\frac{\pdc[2]}{2} \\
&+\frac{1}{2\ell_1}\frac{\Phi_0^2}{4\pi^2}(\pdc[1]-\pxdc[1])^2 \\
& + \frac{1}{2\ell_2}\frac{\Phi_0^2}{4\pi^2}(\pdc[2] - \pxdc[2])^2\\
&+\frac{1}{2L_{1\xi}}\frac{\Phi_0^2}{4\pi^2}(\p[1] - \px[1])^2 +\frac{1}{2L_{2\xi}}\frac{\Phi_0^2}{4\pi^2}(\p[2] - \px[2])^2\\
&+\frac{M_{12}}{\sqrt{L_{1\xi} L_{2\xi}}}\frac{\Phi_0^2}{4\pi^2}(\p[1] - \px[1])(\p[2] - \px[2]),
\end{align*}
where $E_{2\pm1} = \frac{\Phi_0}{2\pi} I_{\pm1}, E_{4\pm3} = \frac{\Phi_0}{2\pi} I_{\pm2}$, $\frac{1}{L_{i\xi}} =\frac{\xi}{L_i}$ and $\xi = (1-m_{12}^2)^{-1}$ . 

Substituting $E_{2\pm1}$ and $E_{4\pm3}$ into $U$ and set $L_1 = L_2 = L$ give:
\begin{align*} 
U = & \frac{\Phi_0}{2\pi} \Big[ I_{+1} \cos\p[1] \cos\frac{\pdc[1]}{2} +  I_{-1} \sin\p[1] \sin\frac{\pdc[1]}{2} \\
& +  I_{+2} \cos\p[2] \cos\frac{\pdc[2]}{2} +  I_{-2} \sin\p[2] \sin\frac{\pdc[2]}{2} \Big] \\
&+\frac{1}{2\ell_1}\frac{\Phi_0^2}{4\pi^2}(\pdc[1]-\pxdc[1])^2 \\
& + \frac{1}{2\ell_2}\frac{\Phi_0^2}{4\pi^2}(\pdc[2] - \pxdc[2])^2\\
&+\frac{\xi}{2L}\frac{\Phi_0^2}{4\pi^2}(\p[1] - \px[1])^2 +\frac{\xi}{2L}\frac{\Phi_0^2}{4\pi^2}(\p[2] - \px[2])^2\\
&+\frac{\xi M_{12}}{\sqrt{L^2}}\frac{\Phi_0^2}{4\pi^2}(\p[1] - \px[1])(\p[2] - \px[2])  ~.
\end{align*}
Simplifying, the expression becomes:
\begin{align*} 
U = & \frac{\Phi_0^2}{4\pi^2L}\Big[ \\
& ~ \beta_1 \cos\p[1] \cos\frac{\pdc[1]}{2} + \delta \beta_1 \sin\p[1] \sin\frac{\pdc[1]}{2} \\
 & + ~ \beta_2 \cos\p[2] \cos\frac{\pdc[2]}{2} + \delta \beta_2 \sin\p[2] \sin\frac{\pdc[2]}{2} \\
& + \frac{\gamma_1}{2}(\pdc[1]-\pxdc[1])^2 + \frac{\gamma_2}{2}(\pdc[2] - \pxdc[2])^2\\
& + \frac{\xi}{2}(\p[1] - \px[1])^2 +\frac{\xi}{2}(\p[2] - \px[2])^2\\
& + m_{12}\xi(\p[1] - \px[1])(\p[2] - \px[2])\Big]
  ~,
\end{align*}
With $U_0 = \frac{\Phi_0^2}{4\pi^2L} $, we can write $U$ as:
\begin{align*} 
U = U_0 U'
~.
\end{align*}

\section{Work Done for Each Substage}
\label{appendix:substage_work_distribution}

External parameters change the shape of the potential, but also induce shifts in the effective zero-point of the potential energy landscape. This can introduce extraneous energetic contributions that are more due to the choice of measuring stick than to information processing. To isolate the inherent information processing costs, we employ a minimum-point offset approach at each time step. This involves subtracting the minimum potential value from the entire landscape, effectively setting it as the zero-point reference. Consequently, the potential of the minimal point remains at a constant reference level throughout the whole cycle. This technique eliminates the influence of the zero-point shifting of the potential landscape without affecting the total work done, as the shift cancels out upon completion of a full cycle. 

Table \ref{table:substage_mean_work} summarizes the mean work for each stage with and without offset. It is hard to see the connection of the value of the work done and the effect of the substage without the offset. After the offset, we can see the pure effect of changing the shape of the potential. Lowering the barrier height (substages 1 and 2) results in negative work. Tilting the potential to move particles from the middle of the sides requires work input (substage 3), while the reverse process releases work (substage 6). Going from conditional tilt back to the four-well potential (substage 4) needs to spend work and the reverse process (substage 5) releases work. Raising the barrier height (substages 7 and 8) yields positive work.

\begin{table}
\begin{tabular}{ | wc{2cm} | wc{2.5cm} | wc{2.5cm}|} 
    \hline
      Substage & $\langle W \rangle (k_BT)$ & $\langle W_{\text{offset}} \rangle (k_BT) $\\
    \hline
    1 & 3.76 & -0.26\\
    \hline
    2 & -49.20 & -1.88\\
    \hline
    3 & -3.18 & 2.51\\
    \hline
    4 & 49.51 & 0.50\\
    \hline
    5 & -49.40 & -0.39\\
    \hline
    6 & 3.14 & -2.55\\
    \hline
    7 & 49.30 & 1.98\\
    \hline
    8 & -3.79 & 0.23\\
    \hline
\end{tabular}
\caption{\label{table:substage_mean_work} Mean work in each substage
    with and without offset. The second column gives the mean work without offset. While the third column shows the mean work with offset.
    }
\end{table}

\begin{figure*}
\includegraphics[scale=0.7]{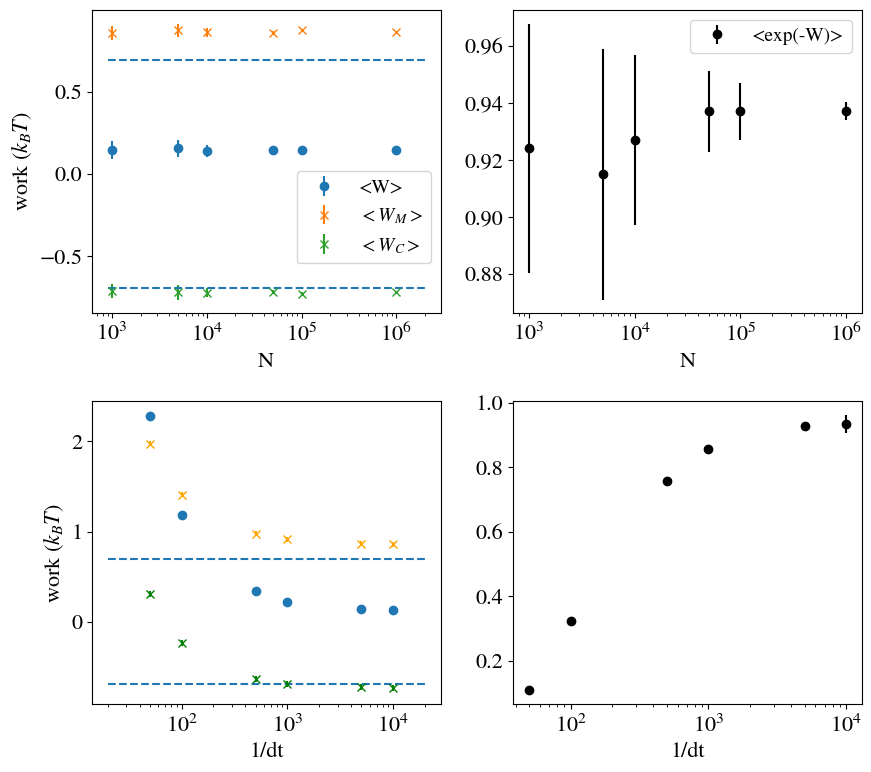}
\caption{Average total work, average work for measurement, average work
    for control, and the Jarzynski term of the coupled QFP for different values of (a) number of samples ($N$) from $10^3$ to $10^6$ and (b) inverse of time step ($1/dt)$ from $50$ to $10,000$ under a symmetric potential. For simulations in (a), the value of $dt$ used was $1/5000$. For simulations in (b), $N = 10,000$ was used. The dash lines indicate the position of $k_BT \ln2$ and $-k_BT \ln2$, the fundamental bounds for measurement work and reset work. The value of $dt$ is more crucial for achieving accurate simulation results compared to the value of $N$.
}
\label{fig:N_and_dt}
\end{figure*}

\begin{figure*}
\centering
 \subfigure{
    \includegraphics[width=\textwidth]{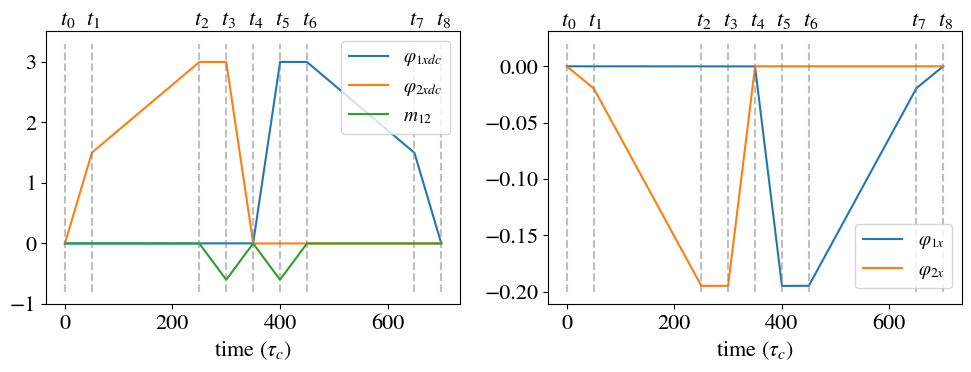}
 }
 \subfigure{
 \begin{tabular}{ | wc{1.7cm} | wc{1.9cm} | wc{1.5cm}| wc{1.5cm} |  wc{1.5cm} | wc{1.7cm}| wc{1.7cm} |} 
         \hline
         substage & duration ($\tau_c$) & $\pxdc[1]$ & $\pxdc[2]$ & $m_{12}$ & $\px[1]$ & $\px[2]$\\
         \hline
         1 ($t_0$ to $t_1$)& 50  & 0  & $0 \rightarrow 1.5$ & 0 & 0 & $0 \rightarrow \px^1$ \\
         \hline
         2 ($t_1$ to $t_2$) & 200  & 0 & $0 \rightarrow 3.0$  & 0 & 0 & $\px^1 \rightarrow \px^2$ \\
         \hline
         3 ($t_2$ to $t_3$) &  50  & 0 & 3.0 & $0 \rightarrow -0.6$  & 0 & $\px^2$ \\
         \hline
         4 ($t_3$ to $t_4$)  & 50  & 0 & $3.0 \rightarrow 0$ & $-0.6 \rightarrow 0$ & 0 & $\px^2 \rightarrow 0$\\
         \hline
         5 ($t_4$ to $t_5$)  & 50  & $0 \rightarrow 3.0$ & 0 & $0 \rightarrow -0.6$ & $0 \rightarrow \px^2$ & 0\\
         \hline
         6 ($t_5$ to $t_6$) & 50  & 3.0 & 0 & $-0.6 \rightarrow 0$ & $\px^2$ & 0\\
         \hline
         7 ($t_6$ to $t_7$) & 200  & $3.0 \rightarrow 1.5$ & 0 & 0 & $\px^2 \rightarrow \px^1$ & 0\\
         \hline
         8 ($t_7$ to $t_8$) & 50  & $1.5 \rightarrow 0$ & 0 & 0 & $\px^1 \rightarrow 0$  & 0\\
         \hline
    \end{tabular}
 }
\caption{Table and the graphs show the values of the external
    parameters $\pxdc[1]$, $\pxdc[2]$, and $m_{12}$, $\px[1]$, and $\px[2]$ at different time of the protocol. For a symmetric potential, $\px^1 = \px^2 = 0$, but for an asymmetric potential with $\delta \beta = 0.170$, $\px^1$ = -0.01947, and $\px^2=-0.1947$. This protocol represents a compensated asymmetry, as described in Sec. \ref{sec:CompAsymm}. The dashed lines in the graphs indicate key times. The table shows the values of the external parameters at different times.}
\label{fig:protocol_values}
\end{figure*}

\section{Simulation Details}
\label{appendix:simulation_detail}

This section gives the details of the numerical simulations and algorithms.

\subsection{Algorithm}

The 4th-order Runge-Kutta method was applied to the deterministic portion and the Euler-Maryama method was used for the stochastic portion of the integration. Figure \ref{fig:N_and_dt} investigates the relationship between the number of samples (N) and the time step (dt) with the average work, average work for measurement, average work for control and the Jarzynski term. The Jarzynski term is the lefthand side term of the Jarzynski equality \cite{PhysRevLett.78.2690}:
\begin{align*}
\left\langle e^{-W/k_BT} \right\rangle = e^{-\Delta F / k_BT}
~.
\end{align*}

The initial and final potential of a complete cycle of the Szilard engine are the same, thus the change in free energy vanishes and the expected value of the Jarzynski term is one. However, due to the dominance of rare events in the average---this Jarzynski term converges much more slowly than the average work \cite{jarzynski2006rare}. A failure of the Jarzynski term to converge to the proper value does not reflect an error in the converged value of average work. The plot reveals that $N$ does not affect simulation results much, but the value of $dt$ is crucial for an accurate result. The results start to converge when $1/dt > 10^3$.

\subsection{Work Done and Fidelity}

The work done by the k-th particle is expressed as:
\begin{align*}
W_k & = \sum^{n}_{i = 0} [U(\vec{x}_k(\tau_i), \tau_{i+1}) - U(\vec{x}_k(\tau_{i}), \tau_{i}) ]\\
  & ~~ -  \sum^{n}_{i = 0} [U_{\text{min}}(\tau_{i+1)} - U_{\text{min}}(\tau_{i)}]
  ~.
\end{align*}
Here, $\tau_i$ is the time at the i-th step, $\vec{x}_k$ is the state of the k-th particle and U is the potential, which is a function of the particle state and time. The upper limit of the summation (n) is the total number of time steps.

The first summation term in the work done is the difference of the potential energy at $\tau_{i+1}$ of the i-th state and that at $\tau_{i+1}$ of the same state. The second summation is the difference of the minimum point of the potential at $\tau_{i+1}$ and $\tau_i$, the offset term mentioned in App. \ref{appendix:substage_work_distribution}. The average work done is the average of the work done of all the particles in the ensemble.

Fidelity evaluates how good the protocol is at bringing particles to their intended location after an operation. An operation having high fidelity causes particles initially located in wells $00$ and $10$ ($10$ and $11$) be each split evenly into quadrant III and IV (I and II).

\subsection{Protocol value}

Figure \ref{fig:protocol_values} shows the values of the five external parameters in a complete cycle. The time is normalized by $\tau_c = \sqrt{L_1C_1}$, where $L_1$ and $C_1$ are the geometric inductance and the capacitance of the QFP. Each row of the table is a substage of the protocol. There are $8$ substages in total. In each substage, the parameters are changed linearly over time. The lowering of barriers (increasing the barriers) is broken into Substage 1 and Substage 2 (Substage 7 and Substage 8) as this gives a lower total mean work done. Note that the value of $\px[1]$ and $\px[2]$ are zero when the Josephson junctions are symmetrical while they are nonzero for the asymmetrical case.

\section{Barrier Escape Rate}
\label{appendix:escape_rate}

The escape rate \cite{PhysRevLett.63.1712, PhysRevB.46.6338, PhysRevB.35.4682} is:
\begin{align*}
    \Gamma_{2D} = \frac{\Omega}{2\pi}
    \exp\left(-\frac{\Delta U}{k_B T}\right),
\end{align*}
where $\Delta U$ is the potential barrier height between the two wells, $\Omega = \omega_{lw} \omega_{tw} / \omega_{ts}$, $\omega_{lw}$ and $\omega_{tw}$ are the longitudinal and transverse plasma frequencies at the bottom of the potential well, and $\omega_{ts}$ is the transverse plasma frequency at the saddle point in between the two potential wells. 

For the $4$-well potential, $\Delta U$ is about $50$ $k_BT$. The saddle points are located at $(0, \pm 2.68)$ and $(\pm 2.68, 0)$. The plasma frequencies at the potential wells and the saddle points can be found by the small-amplitude oscillation expression $\omega = \sqrt{k_{\text{eff}} / m_{\text{eff}}}$, where $m_{\text{eff}}$ is the effective mass and $k_{\text{eff}} = d^2U / d\phi_2^2 = U_{0} / x_c^2 \times d^2U' /d\p[2]^2$. 

Consider the saddle point at $(0, 2.68)$. The longitudinal direction is parallel to the $\p[1]$ axis, and the equation of the potential on the $\p[1] = 0$ surface is $U' = \frac{\p[2]^2}{2} + \beta_2 \cos \p[2] + \beta_1$ and $d^2U' / d\p[2]^2 = 1 - \beta_2 \cos\p[2]$. Therefore, the curvature at the point $(0, 2.68)$ is $d^2U' / d\p[2]^2 |_{\p[2] = 2.68} = 6.44$.
 
Similarly, we find the transverse and longitudinal curvature at the bottom of the potential wells. Consider the point $(2.68, 2.68)$. The potential on the surface $\p[1] = 2.68$ is $U' = \frac{1}{2}\p[2]^2 + \beta_1 \cos\p[2] + \text{constant}$. Via a similar calculation, the curvature at the saddle point is also $6.44$. The curvatures give $\omega_{lw} = \omega_{tw} = \omega_{ts} = 1.27$ GHz. Inserting these results into the escape rate expression, results in $\Gamma_{2D} = 1.50 \times 10^{-43}$ Hz.

\bibliography{references.bib}

\end{document}